\documentclass[]{aastex}
\usepackage{psfig,natbib}
\usepackage{emulateapj5}

\newcommand{\citepeg}[1]{\citep[{e.g.,}][]{#1}}
\newcommand{\citepcf}[1]{\citep[{see}\phantom{}][]{#1}}

\def\lsim{\hbox{ \rlap{\raise 0.425ex\hbox{$<$}}\lower 0.65ex\hbox{$\sim$}}}
\def\gsim{\hbox{ \rlap{\raise 0.425ex\hbox{$>$}}\lower 0.65ex\hbox{$\sim$}}}

\def\arcsec{\hbox{$^{\prime\prime}$}}

\def\ale{\mathrel{\hbox{\rlap{\hbox{\lower4pt\hbox{$\sim$}}}\hbox{$<$}}}}
\def\age{\mathrel{\hbox{\rlap{\hbox{\lower4pt\hbox{$\sim$}}}\hbox{$>$}}}}

\begin{document}

\journalinfo{Published in the {\it Astrophysical Journal Letters},
vol.~572, L45--L49}

\slugcomment{Received 2002 March 21; accepted 2002 May 2; published 2002 May 20
}

\title{Detection of a supernova signature associated with
       GRB 011121$^1$}\phantom{\footnotemark}\footnotetext{Based on
       observations with the NASA/ESA Hubble Space Telescope, obtained
       at the Space Telescope Science Institute, which is operated by
       the Association of Universities for Research in Astronomy,
       Inc.~under NASA contract No.~NAS5-26555.}

\def\cit{2}
\def\mso{3}
\def\vla{4}
\def\uva{5}
\def\uta{6}
\def\col{7}
\def\tap{8}
\def\car{9}
\def\god{10}
\def\nms{11}
\def\ucb{12}
\def\cnr{13}
\def\fer{14}
\def\rom{15}

\author{J. S. Bloom\altaffilmark{\cit},
S. R. Kulkarni\altaffilmark{\cit},
P. A. Price\altaffilmark{\cit,\mso},
D.    Reichart\altaffilmark{\cit},
T. J. Galama\altaffilmark{\cit},
B. P. Schmidt\altaffilmark{\mso},
D. A. Frail\altaffilmark{\cit,\vla},
E.    Berger\altaffilmark{\cit},
P. J. McCarthy\altaffilmark{\car},
R. A. Chevalier\altaffilmark{\uva},
J. C. Wheeler\altaffilmark{\uta},
J. P. Halpern\altaffilmark{\col},
D. W. Fox\altaffilmark{\cit},
S. G. Djorgovski\altaffilmark{\cit},
F. A. Harrison\altaffilmark{\cit},
R.    Sari\altaffilmark{\tap},
T. S. Axelrod\altaffilmark{\mso},
R. A. Kimble\altaffilmark{\god},
J.    Holtzman\altaffilmark{\nms},
K.    Hurley\altaffilmark{\ucb},
\hbox{F.    Frontera\altaffilmark{\cnr, \fer}},
L.    Piro\altaffilmark{\rom},
\&\
E.    Costa\altaffilmark{\rom}}

\altaffiltext{\cit}{Division of Physics, Mathematics and Astronomy,
  105-24, California Institute of Technology, Pasadena, CA 91125}
\altaffiltext{\mso}{Research School of Astronomy \& Astrophysics, 
Mount Stromlo Observatory, via Cotter Rd., Weston Creek 2611, Australia}
\altaffiltext{\vla}{National Radio Astronomy Observatory, Socorro,
NM 87801} 
\altaffiltext{\uva}{Department of Astronomy, University of Virginia, P.O. Box 3818, Charlottesville, VA 22903-0818}
\altaffiltext{\uta}{Astronomy Department, University of Texas, Austin, TX 78712}
\altaffiltext{\col}{Columbia Astrophysics Laboratory, Columbia University, 550 West 120th Street, New York, NY 10027}
\altaffiltext{\tap}{Theoretical Astrophysics 130-33, California Institute
  of Technology, Pasadena, CA 91125}
\altaffiltext{\car}{Carnegie Observatories, 813 Santa Barbara Street, Pasadena, CA 91101}
\altaffiltext{\god}{Laboratory for Astronomy and Solar Physics, NASA Goddard Space Flight Center, Code 681, Greenbelt, MD 20771}
\altaffiltext{\nms}{Department of Astronomy, MSC 4500, New Mexico State University, P.O.~Box 30001, Las Cruces, NM 88003}
\altaffiltext{\ucb}{University of California at Berkeley, Space Sciences Laboratory, Berkeley, CA 94720-7450}
\altaffiltext{\cnr}{Istituto Astrofisica Spaziale and Fisica Cosmica, C.N.R., Via Gobetti, 101, 40129 Bologna, Italy}
\altaffiltext{\fer}{Physics Department, University of Ferrara, Via Paradiso, 12, 44100 Ferrara, Italy}
\altaffiltext{\rom}{Istituto Astrofisica Spaziale, C.N.R., Area di Tor Vergata, Via Fosso del Cavaliere 100, 00133 Roma, Italy}

\begin{abstract}
  Using observations from an extensive monitoring campaign with the
  {\it Hubble Space Telescope} we present the detection of an
  intermediate time flux excess which is redder in color relative to
  the afterglow of GRB 011121, currently distinguished as the
  gamma-ray burst with the lowest known redshift.  The red ``bump,''
  which exhibits a spectral roll-over at $\sim$7200 \AA, is well
  described by a redshifted Type Ic supernova that occurred
  approximately at the same time as the gamma-ray burst event.  The
  inferred luminosity is about half that of the bright supernova
  1998bw. These results serve as compelling evidence for a massive
  star origin of long-duration gamma-ray bursts. Models that posit a
  supernova explosion weeks to months preceding the gamma-ray burst
  event are excluded by these observations. Finally, we discuss the
  relationship between spherical core-collapse supernovae and
  gamma-ray bursts.
\end{abstract}

\keywords{supernovae: general --- gamma rays:
bursts --- supernovae: individual (SN 1998bw)}

\section{Introduction}

Two broad classes of long-duration gamma-ray burst (GRB) progenitors
have survived scrutiny in the afterglow era: the coalescence of
compact binaries (see \citealt{fwh99} for review) and massive stars
\citep{woo93}. More exotic explanations \citepeg{pac88,car92,der96}
fail to reproduce the observed redshift distribution, detection of
transient X-ray lines, and/or the distribution of GRBs about host
galaxies.

In the latter viable scenario, the so-called ``collapsar'' model
\citep{woo93,mw99,han99}, the core of a massive star collapses to a
compact stellar object (such as a black hole or magnetar) which then
powers the GRB while the rest of the star explodes.  We expect to see
two unique signatures in this scenario: a rich circumburst medium fed
by the mass-loss wind of the progenitor \citep{cl99} and an underlying
supernova (SN). Despite extensive broadband modeling of afterglows,
unambiguous signatures for a wind-stratified circumburst media have
not been seen \citepeg{fks+00,bdf+01}.

There has, however, been been tantalizing evidence for an underlying
SN.  The first association of a cosmologically distant GRB with the
death of a massive star was found for GRB 980326, where a clear excess
of emission was observed, over and above the rapidly decaying
afterglow component. This late-time ``bump'' was interpreted as
arising from an underlying SN \citep{bkd+99} since, unlike the
afterglow, the bump was very red.  GRB 970228, also with an
intermediate-time bump and characteristic SN spectral rollover, is
another good candidate \citep{rei99,gtv+00}.

Suggestions of intermediate-time bumps in GRB light curves have since
been put forth for a number of other GRBs
\citep{lcg+01,svb+00,fvh+00,bhj+01,csg+01,sok01,ddr02}. Most of these
results are tentative or suspect with the SN inferences relying on a
few mildly deviant photometric points in the afterglow light curve. Even
if some of the bumps are real, a number of other explanations for
the physical origin of such bump have been advanced: for example, dust
echoes \citep{eb00,rei01b}, shock interaction with circumburst density
discontinuities \citepeg{rdm+01}, and thermal re-emission of the
afterglow light \citep{wd00}. To definitively distinguish between the
SN hypothesis and these alternatives, detailed spectroscopic
and multi-color light curve observations of intermediate-time bumps are
required. 

It is against this background that we initiated a program with the
{\it Hubble Space Telescope} (HST) to sample afterglow light curves at
intermediate and late-times. The principal attractions of HST
are the photometric stability and high angular resolution. These are
essential in separating the afterglow from the host galaxy and in
reconstructing afterglow colors.

On theoretical grounds, if the collapsar picture is true, then we
expect to see a Type Ib/Ic SN \citep{woo93}.  In the first month,
core-collapsed supernova spectra are essentially characterized by a
blackbody (with a spectral peak near $\sim$5000 \AA) modified by broad
metal-line absorption and a strong flux suppression blueward of
$\sim4000$\, \AA\ in the restframe.  For GRBs with low redshifts, $z
\ale 1$, the effect of this blue absorption blanketing is a source
with an apparent red spectrum at observer-frame optical wavelengths;
at higher redshifts, any supernova signature is highly suppressed. For
low redshift GRBs, intermediate-time follow-up are, then, amenable to
observations with the Wide Field Planetary Camera 2 (WFPC2). In this
{\it Letter} we report on WFPC2 multi-color photometry of GRB 011121
($z=0.36$; \citealt{igsw01}) and elsewhere we report on observations
of GRB 010921 ($z=0.451$; \citealt{psk02}).  In a companion paper
(\citealt{price02a}; hereafter Paper II), we report a multi-wavelength
(radio, optical and NIR) modeling of the afterglow.

\section{Observations and Reductions}

\subsection{Detection of GRB 011121 and the afterglow}

On 2001 November 21.7828 UT, the bright GRB 011121 was detected and
localized by {\it BeppoSAX} to a 5-arcmin radius uncertainty
\citep{pir+01a}.  Subsequent observations of the error circle refined
by the IPN and {\it BeppoSAX} (see Paper II) revealed a fading optical
transient (OT) \citep{wsg01,sgw01}.  Spectroscopic observations with
the Magellan 6.5-m telescope revealed redshifted emission lines at the
OT position ($z=0.36$), indicative of a bright, star-forming host
galaxy of GRB 011121 \citep{igsw01}.

\subsection{HST Observations and reductions}

For all the HST visits, the OT and its underlying host were placed
near the serial readout register of WF chip 3 (position {\sc WFALL})
to minimize the effect of charge transfer (in)efficiency (CTE).  The
data were pre-processed with the best bias, dark, and flat-field
calibrations available at the time of retrieval from the archive
(``on--the--fly'' calibration).  We combined all of the images in each
filter, dithered by sub-pixel offsets, using the standard IRAF/DITHER2
package to remove cosmic rays and produce a better sampled final image
in each filter.  An image of the region surrounding the transient is
shown in figure 1. The point source was detected at
better than 20 $\sigma$ in epochs one, two and three in all filters,
and better than 5 $\sigma$ in epoch four.

\vskip 0.4cm
\begin{minipage}[b]{8.5cm}
\centerline{\psfig{file=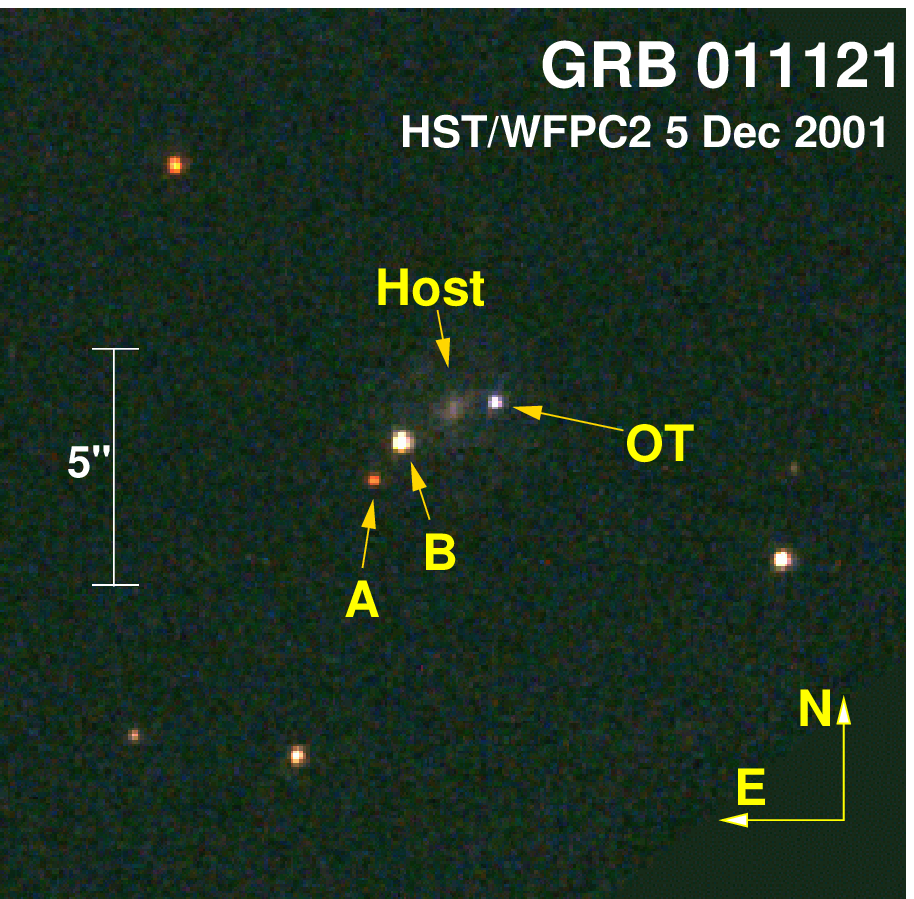,width=8.2cm,angle=0}}
\vskip 0.2cm 
\figcaption[]{{\it Hubble Space Telescope} image of the field of GRB
011121 on 2001 December 4--6 UT.  This false-color image was
constructed by registering the final drizzled images in the F555W
(blue), F702W (green) and F814W (red) filters.  The optical transient
(OT) is clearly resolved from the host galaxy and resides in the
outskirts of the morphologically smooth host galaxy. Following the
astrometric methodology outlined in \citet{bkd02}, we find that the
transient is offset from the host galaxy (883 $\pm$ 7) mas west, (86
$\pm$ 13) mas north.  The projected offset is (4.805 $\pm$ 0.035) kpc,
almost exactly at the host half-light radius. Sources ``A'' and ``B''
are non-variable point sources that appear more red than the OT and
are thus probably foreground stars.\vskip 0.2cm}
\label{fig:1121im}
\end{minipage}

Given the proximity of the OT to its host galaxy, the final HST images
were photometered using the {\sc IRAF/DAOPHOT} package which
implements PSF-fitting photometry on point-sources \citep{ste87}.  The
PSF local to the OT was modeled with {\sc PSTSELECT} and {\sc PSF}
using at least 15 isolated stars detected in the WF chip 3 with an
adaptive kernel to account for PSF variations across the image ({\sc
VARORDER = 1}).  The resulting photometry, reported in Table
\ref{tab:hst-mag}, was obtained by finding the flux in an 0\arcsec .5
radius using a PSF fit.  We corrected the observed countrate using the
formulation for CTE correction in \citet{dol00} with the most
up-to-date parameters\footnotemark\footnotetext{See {\tt
http://www.noao.edu/staff/dolphin/wfpc2\_calib/}}; such corrections,
computed for each individual exposure, were never larger than 8\%
(typically 4\%) for a final drizzled image. We estimated the
uncertainty in the CTE correction, which is dependent upon source
flux, sky background, and chip position, by computing the scatter in
the CTE corrections for each of the images that were used to produce
the final image.  The magnitudes reported in the standard bandpass
filters in Table \ref{tab:hst-mag} were found using the
\citeauthor{dol00} prescription.

\section{Results}

In figure~2 we plot the measured fluxes from our four HST epochs in
the F555W, F702W, F814W and F850LP filters.  We also plot measurements
made at earlier times (0.5 d $< t <$ 3 d) with ground-based telescopes
and reported in the literature.  These magnitudes were converted to
fluxes using the zero-points of \citet{fsi95} and plotted in the
appropriate HST filters.

\begin{figure*}[tbh]
\centerline{\psfig{file=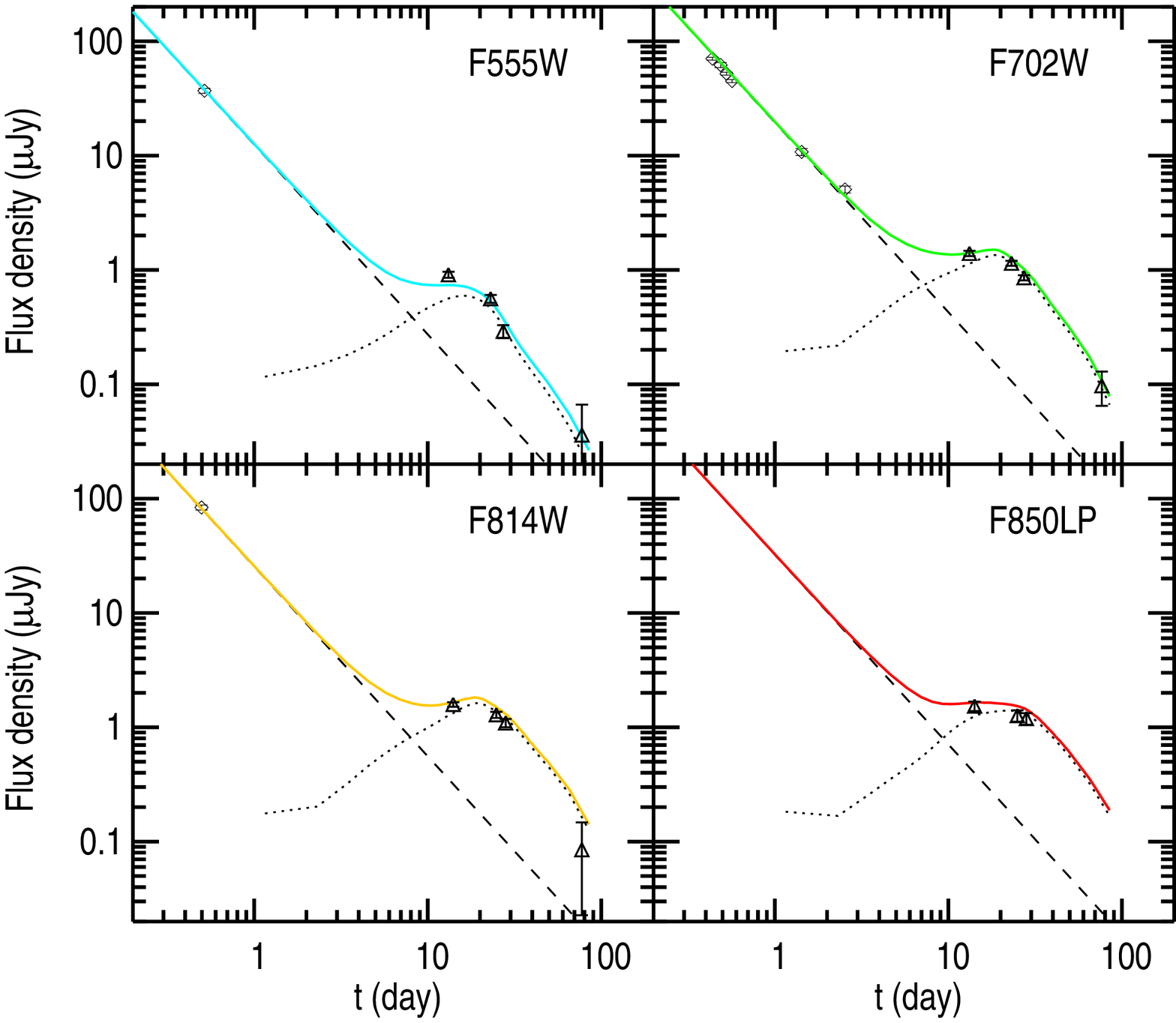,width=5.5in,angle=0}}
\figcaption[]{Light-curves of the afterglow and the intermediate-time red
bump of GRB~011121.  The triangles are our HST photometry in the
F555W, F702W, F814W and F850LP filters (all corrected for the
estimated contribution from the host galaxy), and the diamonds are
ground-based measurements from the literature \citep{obs+01,kw01}.
The dashed line is our fit to the optical afterglow (see Paper II),
the dotted line is the expected flux from the template SN at the
redshift of GRB~011121, with foreground extinction applied and dimmed
by 55\% to approximately fit the data, and the solid line is the sum
of the afterglow and SN components. Corrections for color effects
between the ground-based filters and the HST filters were taken to be
negligible for the purpose of this exercise.}
\label{fig:lc}
\end{figure*}

\begin{figure*}[tbh]
\centerline{\psfig{file=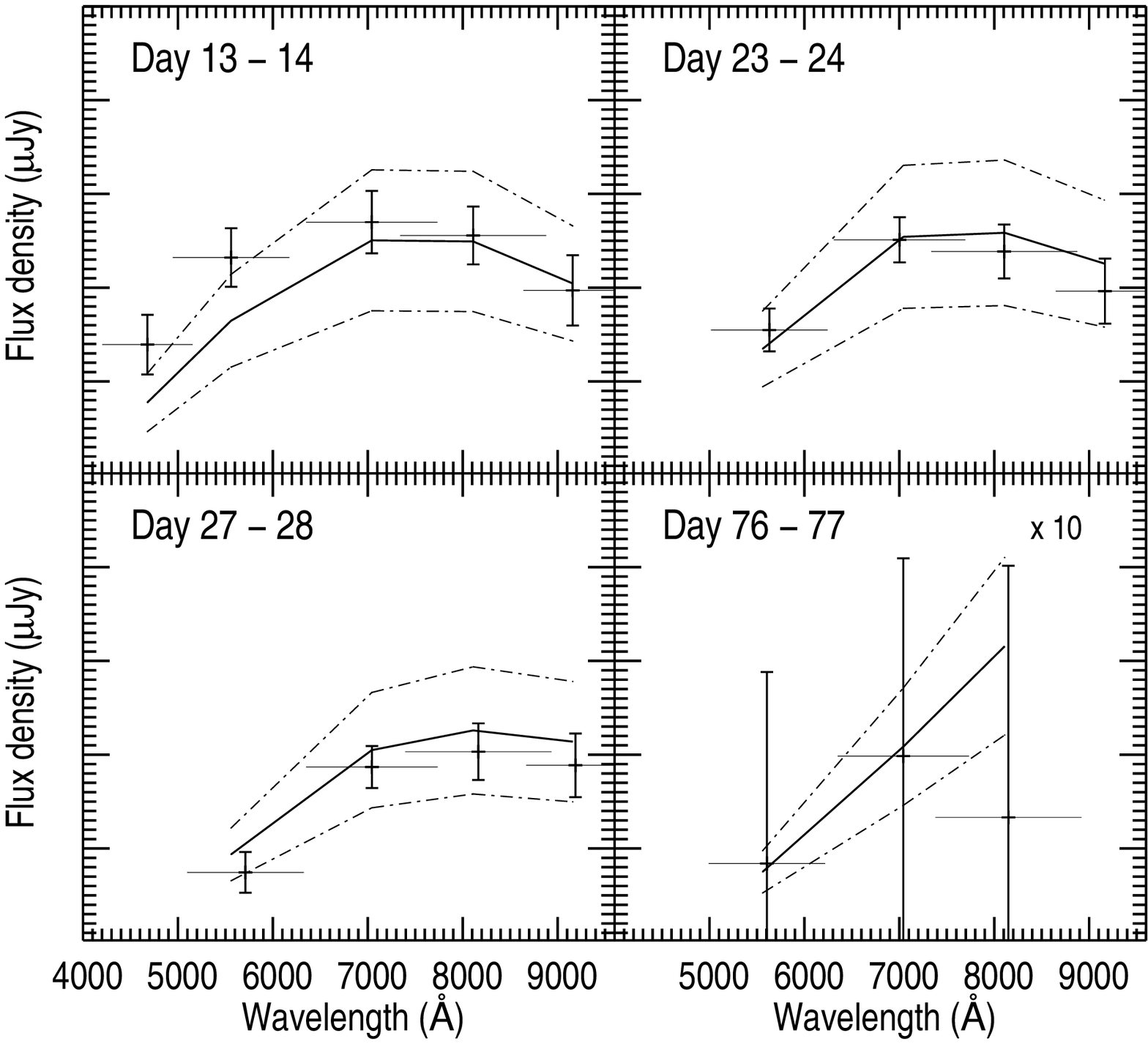,width=5.5in,angle=0}} 
\figcaption[The spectral flux distributions of the red bump at the time
of the four HST epochs]{The spectral flux distributions of the red
bump at the time of the four HST epochs. The fluxes are dereddened
using $A_V = 1.16$ mag.  Spectral evolution, and more important, a
turn-over in the spectra of the first three epochs, are clearly seen.
The peak of the turn-over (around $7200$ \AA) corresponds to a peak in
the red bump spectrum at $\sim$5300 \AA. For comparison, we show a
template broadband SN spectra (a dimmed version of SN 1998bw; solid
curve) as it would appear at the redshift of GRB 011121 and the
associated 2 $\sigma$ errors (see text).  The vertical error bars on
the red bump reflect the 1 $\sigma$ statistical uncertainty flux from
only the red bump.  There are large ($\sim$1 mag) systematic
uncertainties (e.g., Galactic reddening, relative distance moduli
between SN 1998bw and GRB 011121) in both the data and the model;
these are suppressed for clarity.}
\label{fig:sfd}
\end{figure*}
Corrections
for color effects between the ground-based filters and HST filters
were taken to be negligible for the purpose of this exercise.

The estimated contribution from the afterglow is heavily weighted by
the available data: our ground-based data (and those reported in the
literature so far) are primarily at early times. Roughly, over the
first week, the afterglow exhibits a simple power law decay. The
afterglow contribution derived from our NIR data and optical data from
the literature (see Paper II) is shown by the dashed line in each
panel. No afterglow light curve breaks (e.g., from jetting) were assumed.

\citet{ghj+02} drew attention to an excess of flux (in $R$-band), at a
time 13~days after the GRB, with respect to that expected from the
power-law extrapolation of early-time afterglow emission; they
suggested the excess to arise from an underlying SN. As can vividly be
seen from our multi-color data, the excess is seen in all bands and
over several epochs.

We used the light curve and spectra\footnotemark\footnotetext{Spectra
were obtained through the Online Supernova Spectrum Archive (SUSPECT)
at {\tt http://tor.nhn.ou.edu/$\sim$suspect/index.html}.} of the
well-studied Type Ic supernova SN~1998bw \citep{gvv+98,ms99} to create
a comparison template broad-band light curve of a Type Ic supernova at
redshift $z = 0.36$. Specifically, the spectra of SN 1998bw were used
to compute the $K$-corrections between observed photometric bands of
1998bw and {\it HST} bandpasses (following \citealt{kgp96} and
\citealt{ssp+98}).  A flat $\Lambda$ cosmology with $H_0 = 65$ km
s$^{-1}$ Mpc$^{-1}$ and $\Omega_M=0.3$ was assumed and we took the
Galactic foreground extinction to SN 1998bw of $A_V=0.19$ mag
\citep{gvv+98}.

Since dimmer Ic SNe tend to peak earlier and decay more quickly (see
fig.~1 of \citealt{imn+98}), much in the same way that SN Ia do, we
coupled the flux scaling of SN 1998bw with time scaling in a method
analogous to the ``stretch'' method for SN Ia distances
\citep{pgg+97}.  To do so, we fit an empirical relation between 1998bw
and 1994I to determine the flux-time scaling. We estimate that a
1998bw-like SN that is dimmed by 55\% (see below), would peak and
decay about 17\% faster than 1998bw itself. Some deviations from our
simple one-parameter template are apparent, particularly in the F555W
band and at late-times.

In figure 3, we plot the spectral flux distributions (SFDs) of the
intermediate-time bump at the four HST epochs. A clear turn-over in
the spectra in the first 3 epochs is seen at about 7200 \AA. The solid
curve is the SFD of SN~1998bw transformed as described above with the
associated 2-$\sigma$ errors. Bearing in mind that there are large
systematic uncertainties in the template (i.e., the relative distance
moduli between SN 1998bw and GRB 011121) and in the re-construction of
the red bump itself (i.e., the Galactic extinction toward GRB~011121
and the contribution from the afterglow in the early epochs), the
consistency between the measurements and the SN is reasonable. We
consider the differences, particularly the bluer bands in epoch one,
to be relatively minor compared with the overall agreement. This
statement is made in light of the large observed spectral diversity of
Type Ib/Ic SNe (see, for example, figure 1 of \citealt{mdm+02}).

\section{Discussion and Conclusions}
\label{sec:bw-dis}

We have presented unambiguous evidence for a red, transient excess
above the extrapolated light curve of the afterglow of GRB~011121.  We
suggest that the light curve and spectral flux distribution of this
excess appears to be well represented by a bright SN.  While we have
not yet explicitly compared the observations to the expectations of
alternative suggestions for the source of emission (dust echoes,
thermal re-emission from dust, etc.), the simplicity of the SN
interpretation---requiring only a (physically motivated) adjustment in
brightness---is a compelling (i.e., Occam's Razor) argument to accept
our hypothesis.  Given that the red bump detections in a number of
other GRBs occur on a similar timescale as in GRB 011121, any model
for these red bumps should have a natural timescale for peak of $\sim
20 (1 + z)$ day; in our opinion, the other known possibilities do not
have such a natural timescale as compared with the SN
hypothesis. Indeed, if our SN hypothesis is correct, then the flux
should decline as an exponential from epoch four onward.  The ultimate
confirmation of the supernova hypothesis is a spectrum which should
show characteristic broad metal-line absorption of the expanding
ejecta (from, e.g., Ca II, Ti II, Fe II).

We used a simplistic empirical brightness--time stretch relation to
transform 1998bw, showing good agreement between the observations and
the data. If we neglect the time-stretching and only dim the 1998bw
template, then the data also appear to match the template reasonably
well, however, the discrepancies in the bluer bands become somewhat
larger and the flux ratios between epochs are slightly more
mismatched.  The agreement improves if we shift the time of the
supernova to be about $\sim$3--5 days (restframe) before the GRB time.
Occurrence times more than about ten days (restframe) before the GRB
can be ruled out. This observation, then, excludes the original
``supranova'' idea \citep{vs98}, that posited a supernova would
precede a GRB by several years (see eq.~[1] of \citealt{vs98}).
Modified supranova scenarios that would allow for any time delay
between the GRB and the accompanying SN, albeit {\it ad hoc}, are
still consistent with the data presented
herein\footnotemark\footnotetext{The explosion date of even very
well-studied supernovae, such as 1998bw, cannot be determined via
light curves to better than about 3 days (e.g.,
\citealt{imn+98}). This implies that future photometric studies might
not be equipped to distinguish between contemporaneous SN/GRB events
and small delay scenarios.}.

Regardless of the timing between the SN explosion and the GRB event
(constrained to be less than about 10 days apart), the bigger picture
we advocate is that GRB 011121 resulted from an explosive death of a
massive star.  This conclusion is independently supported by the
inference, from afterglow observations of GRB 011121 (Paper II), of a
wind-stratified circumburst medium.

The next phase of inquiry is to understand the details of the
explosion and also to pin down the progenitor population. A large
diversity in any accompanying SN component of GRBs is expected from
both a consideration of SNe themselves and the explosion mechanism.
The three main physical parameters of a Type Ib/Ic SN are the total
explosive energy, the mass of the ejecta, and the amount of Nickel
synthesized by the explosion ($M_{\rm Ni}$). The peak luminosity and
time to peak are roughly determined by the first two whereas the
exponential tail is related to $M_{\rm Ni}$.  Ordinary Ib/Ic SNe
appear to show a wide dispersion in the peak luminosity
\citep{imn+98}. There is little {\it ab initio} understanding of this
diversity (other than shifting the blame to dispersion in the three
parameters discussed above).

It is now generally accepted that GRBs are not spherical explosions
and are, as such, usually modeled as a jetted outflow. \citet{fks+01}
model the afterglow of GRBs and have presented a compilation of
opening angles, $\theta$, ranging from less than a degree to 30
degrees and a median of 4 degrees. If GRBs have such strong
collimation then it is not reasonable to assume that the explosion,
which explodes the star, will be spherical. We must be prepared to
accept that the SN explosion is extremely asymmetric and thus even a
richer diversity in the light curves. This expected diversity may
account for both the scale factor difference between the SN component
seen here and in SN 1998bw seen in figure 3.  Indeed,
there has been a significant discussion in the literature as to the
degree which the central engine in GRBs will affect the overall
explosion of the star \citep{woo93,kho99,mw99,hww99,mwh01}.  These
models currently have focused primarily on the hydrodynamics and lack
the radiative modeling necessary to compare observations to the
models.

Clearly, the next step is to obtain spectroscopy (and perhaps even
spectropolarimetry) and to use observations to obtain a rough measure
of the three-dimensional velocity field and geometry of the debris.
As shown by GRB 011121 the SN component is bright enough to undertake
observations with the largest ground-based telescopes.

We end by noting the following curious point.  The total energy yield
of a GRB is usually estimated from the gamma-ray fluence and an
estimate of $\theta$ \citepcf{fks+01}. Alternatively, the energy in the
afterglow is used \citepeg{pkpp01}. However, for GRB~011121, the
energy in the SN component (scaling from the well-studied SN~1998bw)
is likely to be comparable or even larger than that seen in the burst
or the afterglow. In view of this, the apparent constancy of the
$\gamma$-ray energy release is even more mysterious.

\acknowledgments

We thank S.~Woosley, who, as referee, provided helpful insights toward
the improvement of this work. A.~MacFadyen and E.~Ramirez-Ruiz are
acknowledged for their constructive comments on the paper. J.~S.~B.~is
a Fannie and John Hertz Foundation Fellow. F.~A.~H.~acknowledges
support from a Presidential Early Career award. S.~R.~K.~and
S.~G.~D.~thank the NSF for support. R.~S.~is grateful for support from
a NASA ATP grant. R.~S.~and T.~J.~G.~acknowledge support from the
Sherman Fairchild Foundation.  J.~C.~W.~acknowledges support from NASA grant
NAG59302.  KH is grateful for Ulysses support under JPL Contract
958056, and for IPN support under NASA Grants FDNAG 5-11451 and NAG
5-17100. Support for Proposal number HST-GO-09180.01-A was provided by
NASA through a grant from Space Telescope Science Institute, which is
operated by the Association of Universities for Research in Astronomy,
Incorporated, under NASA Contract NAS5-26555.

\begin{deluxetable}{lccccr}
\singlespace
\tablecolumns{6} 
\tablewidth{0in}
\tablecaption{Log of HST Imaging of the optical transient of GRB 011121\label{tab:hst-mag}}
\tabletypesize{\footnotesize}
\tablehead{
\colhead{Filter} & \colhead{$\Delta t$\tablenotemark{a}}  & \colhead{Int.} & \colhead{$\lambda_{\rm eff}$\tablenotemark{b}} &
\colhead{$f_\nu(\lambda_{\rm eff})$\tablenotemark{b}} & \colhead{Vega} \\
\colhead{ } & \colhead{ } & \colhead{Time} & \colhead{ } &  \colhead{ } & \colhead{Magnitude\tablenotemark{c}} \\
\colhead{ } & \colhead{(days)} & \colhead{(sec)} & \colhead{(\AA)} & \colhead{($\mu$Jy)} & \colhead{(mag)}}

\startdata
\sidehead{Epoch 1}
F450W     & 13.09   & 1600  & 4678.52   & 0.551 $\pm$ 0.037    & $B$ = 24.867 $\pm$  0.073   \\  
F555W     & 13.16   & 1600  & 5560.05   & 0.996 $\pm$ 0.049    & $V$ = 23.871 $\pm$  0.056    \\  
F702W     & 13.23   & 1600  & 7042.48   & 1.522 $\pm$ 0.072    & $R$ = 23.211 $\pm$  0.054    \\  
F814W     & 14.02   & 1600  & 8110.44   & 1.793 $\pm$ 0.042    & $I$ = 22.772 $\pm$  0.032    \\  
F850LP    & 14.15   & 1600  & 9159.21   & 1.975 $\pm$ 0.103    &     \\  
\sidehead{Epoch 2}
F555W     & 23.03   & 1600  & 5630.50   & 0.647 $\pm$ 0.035    & $V$ = 24.400 $\pm$  0.061    \\  
F702W     & 23.09   & 1600  & 7002.71   & 1.271 $\pm$ 0.051    & $R$ = 23.382 $\pm$  0.048    \\  
F814W     & 24.83   & 1600  & 8105.05   & 1.495 $\pm$ 0.053    & $I$ = 22.982 $\pm$  0.043    \\  
F850LP    & 24.96   & 1600  & 9166.39   & 1.708 $\pm$ 0.100    &     \\  
\sidehead{Epoch 3}
F555W     & 27.24   & 1600  & 5711.00   & 0.378 $\pm$ 0.027    & $V$ = 25.071 $\pm$  0.076    \\  
F702W     & 27.30   & 1600  & 7043.85   & 0.981 $\pm$ 0.036    & $R$ = 23.697 $\pm$  0.044    \\  
F814W     & 28.10   & 1600  & 8164.90   & 1.301 $\pm$ 0.070    & $I$ = 23.157 $\pm$  0.061    \\  
F850LP    & 28.16   & 1600  & 9188.39   & 1.635 $\pm$ 0.092    &     \\
\sidehead{Epoch 4}
F555W     & 77.33   & 2100  & 5604.61   & 0.123 $\pm$ 0.014    & $V$ = 26.173 $\pm$  0.118    \\
F702W     & 76.58   & 4100  & 7042.09   & 0.224 $\pm$ 0.019    & $R$ = 25.264 $\pm$  0.092    \\ 
F814W     & 77.25   & 2000  & 8149.18   & 0.294 $\pm$ 0.020    & $I$ = 24.762 $\pm$  0.073   

\enddata

\tablenotetext{a}{Mean time since GRB trigger on 21.7828 Nov 2001 UT.}
\tablenotetext{b}{In the fourth column, the effective wavelength of
the filter based upon the observed spectral flux distribution of the
transient at the given epoch. In the fifth column, the flux is given
at this effective wavelength in an 0\arcsec.5 radius. The observed
count rate, corrected for CTE effects, was converted to flux using the
{\sc IRAF/SYNPHOT} package. An input spectrum with $f_\nu =$ constant
was first assumed. Then approximate spectral indices between each
filter were computed and then used to re-compute the flux and the
effective wavelength of the filters.  This bootstrapping converged
after a few iterations. The HST photometry contains an unknown but
small contribution from the host galaxy at the OT location. We
attempted to estimate the contamination of the host at the transient
position by measuring the host flux in several apertures at
approximate isophotal levels to the OT position. We estimate the
contribution of the host galaxy to be $f_\nu(F450W) = (0.098 \pm
0.039)~\mu$Jy, $f_\nu(F555W) = (0.087 \pm 0.027)~\mu$Jy, $f_\nu(F702W)
= (0.127 \pm 0.026)~\mu$Jy, $f_\nu(F814W) = (0.209 \pm 0.059)~\mu$Jy,
and $f_\nu(F850LP) = (0.444 \pm 0.103)~\mu$Jy. To correct these
numbers to ``infinite aperture'' multiply the fluxes by 1.096
\citep{hbc+95}. These fluxes have not been corrected for Galactic or
host extinction. } \tablenotetext{c}{Tabulated brightnesses in the
Vega magnitude system ($B_{\rm Vega} = 0.02$ mag, $V_{\rm Vega} =
0.03$ mag, $R_{\rm Vega} = 0.039$ mag, $I_{\rm Vega} = 0.035$ mag;
\citealt{hbc+95}). Subtract 0.1 mag from these values to get the
infinite aperture brightness. These magnitudes have not been corrected
for Galactic or host extinction.}
\end{deluxetable}

\end{document}